\documentclass[a4paper,twoside]{article}

\newcommand{\affil}[1]{$^{\rm #1}$}

\def\simlt{\lower.5ex\hbox{$\; \buildrel < \over \sim \;$}}
\def\simgt{\lower.5ex\hbox{$\; \buildrel > \over \sim \;$}}
\def\arcdeg{\hbox{$^\circ$}}
\def\arcmin{\hbox{$^\prime$}}

\usepackage[authoryear]{natbib}
\bibpunct{(}{)}{;}{a}{}{,}

\usepackage{graphicx}

\date{} 

\title{\large\bf\flushleft 
Paper Productivity of Ground-based Large Optical Telescopes from 2000 to 2009}

\author{\parbox{\textwidth}{\flushleft
\vspace{-0.5cm}
{\it Sang Chul Kim\affil{A,B}}\\
\vspace{0.4cm}
{\small \affil{A}\,Korea Astronomy and Space Science Institute,
  Daejeon 305-348, Republic of Korea}\\
{\small \affil{B}\,Email: sckim@kasi.re.kr}}}

\begin{document}

\twocolumn[
\begin{changemargin}{.8cm}{.5cm}
\begin{minipage}{.9\textwidth}
\vspace{-1cm}
\maketitle


{\small{\bf Abstract:} \\
We present an analysis of the scientific (``refereed'') paper productivity
  of the current largest (diameter $>8$ m) ground-based optical(-infrared) telescopes
  during the ten year period from 2000 to 2009.
The telescopes for which we have gathered and analysed 
  the scientific publication data are
  the two 10 m Keck telescopes, the four 8.2 m Very Large Telescopes (VLT),
  the two 8.1 m Gemini telescopes, the 8.2 m Subaru telescope, and
  the 9.2 m Hobby-Eberly Telescope (HET).
We have analysed the rate of papers published in various astronomical journals
  produced by using these telescopes.
While the total numbers of papers from these observatories
  are largest for the VLT followed by Keck, Gemini, Subaru, and HET,
  the number of papers produced by each component of the telescopes
  are largest for Keck followed by VLT, Subaru, Gemini, and HET.
In 2009, each telescope of the Keck, VLT, Gemini, Subaru, and HET observatories
  produced 135, 109, 93, 107, and 5 refereed papers, respectively.
We have shown that each telescope of the Keck, VLT, Gemini, and Subaru observatories
  is producing $2.1 \pm 0.9$ {\it Nature} and {\it Science} papers annually and
  the rate of these papers among all the refereed papers
  produced by using that telescope is $1.7 \pm 0.8$ \%.
Extending this relation, we propose that this ratio of the number of 
  {\it Nature} and {\it Science} papers over the number of whole
  refereed papers that will be produced by future extremely large telescopes (ELTs)
  will be remained similar.
From the comparison of the publication trends of the above telescopes, we suggest
  that (i) having more than one telescope of the same kind
  at the same location and
  (ii) increasing the number of instruments available at the telescope
  are good ways to maximize the paper productivity. }


\medskip
\medskip
\end{minipage}
\end{changemargin}
]
\small

\section{Introduction}
\label{sect:intr}

Astronomy is a science driven by 
  discovery\footnote{http://www.gmto.org/sciencecase/GMT-ID-01404-GMT\_Science\_Case.pdf}, 
  and the essential components in the astronomical discoveries 
  are telescopes and instruments.
Since 1990, the world's largest optical telescopes with 8 to 10 m in diameter
  have appeared and gave birth to new innovations in astronomy.
After 10 -- 20 years of use by the largest optical telescopes,
  the necessity for larger telescope has increased and 
  we are witnessing the development of 
  25 -- 42 m extremely large telescopes (ELTs).

Table \ref{tab1} lists the current largest, ground-based optical telescopes 
(diameter D $\ge$8 m).
There are five, more active telescopes mainly built in the 20th century :
USA's two 10 m Keck telescopes,
  four 8.2 m Very Large Telescope (VLT) of the European Southern Observatory (ESO),
  two 8.1 m Gemini telescopes of the consortium of 
  USA, UK, Canada, Chile, Australia, Brazil, and Argentina, 
  8.2 m Subaru telescope of Japan, and
  9.2 m Hobby-Eberly Telescope (HET).
The other three telescopes at the lower part of Table \ref{tab1}
  are built later in the 21st century, and
  are in the early part of their operations, 
  which are the 10.4 m Gran Telescopio Canarias (GTC),
  the 10 m South African Large Telescope (SALT), and
  the two 8.4 m (on a single mount) Large Binocular Telescope (LBT).

\begin{table*} 
\scriptsize{
\begin{center}
\caption{Current Ground-based Large Optical Telescopes (diameter $\ge$8 m)}\label{tab1}
\begin{tabular}{cccc}
\hline\hline
{Telescope} & {Diameter} & {Partner Country/Institute} & {Operation} \\
&&& {Start Year} \\
\hline
Keck & 10 m (36 $\times$ 1.8 m segments) $\times$ 2 & USA (Caltech, University of
  California) & 1993, 1996 \\
VLT$^{\rm a}$ & 8.2 m $\times$ 4 & ESO members (Austria,
  Belgium, Czech Republic, & 1998, 1999, \\
 & & Denmark, Finland, France, Germany, Italy, Netherlands, & 2000, 2000 \\
 & & Portugal, Spain, Sweden, Switzerland, United Kingdom) & \\
Gemini & 8.1 m $\times$  2 (N, S hemispheres) & USA (48\%), UK (24\%),
  Canada (14\%), Chile (5\%), & 1999, 2000 \\
 & & Australia (5\%), Brazil (2\%), Argentina (2\%) & \\
Subaru & 8.2 m & Japan & 1999 \\
HET$^{\rm b}$ & 9.2 m (91 segments) &
   USA (90\%), Germany (10\%) & 1999 \\
\hline
GTC$^{\rm c}$ & 10.4 m (36 segments) & Spain (90\%),
  University of Florida (5\%), Mexico (5\%) & 2008 \\
SALT$^{\rm d}$ & 10 m (91 segments) &
  Republic of South Africa (35\%), Poland, HET, USA, & \\
 & & Germany, New Zealand, 
  UK, India & 2005 \\
LBT$^{\rm e}$ & 8.4 m $\times$ 2 (single mount) & 
  USA (50\%), Italy (25\%), Germany (25\%) &
  2008 \\
\hline
\end{tabular}
\medskip\\
\end{center}
$^{\rm a}$ Very Large Telescope \\
$^{\rm b}$ Hobby-Eberly Telescope \\
$^{\rm c}$ Gran Telescopio Canarias \\
$^{\rm d}$ South African Large Telescope \\
$^{\rm e}$ Large Binocular Telescope \\
} 
\end{table*}

Table \ref{tab2} lists the three planned ELT projects,
  of which telescope diameters are larger than 20 m.
When these state-of-the-art telescopes are completed,
  how much impact will they bring to the astronomical research?
One of the methods to answer this question is to scrutinize the impacts
  the current largest telescopes have brought since their completions.
The analysis of the major scientific papers published by using these telescopes
  would be the essential part in this area,
  which will be shown in this paper.

\begin{table*} 
{\footnotesize 
\begin{center}
\caption{Extremely Large Telescope (ELT) Projects in Progress\label{tab2}}
\begin{tabular}{ccc}
\hline\hline
{Telescope} & {Diameter} & {Partner Country/Institute} \\
\hline
Giant Magellan Telescope (GMT) & 25 m (7 $\times$ 8.4 m segments) & USA, Korea, Australia \\
Thirty Meter Telescope (TMT) & 30 m (492 $\times$ 1.4 m segments) & USA, Canada$^{\rm a}$ \\
European Extremely Large Telescope (E-ELT) & 42 m (984 $\times$ 1.4 m segments) & ESO \\
\hline
\end{tabular}
\medskip\\
\end{center}
$^{\rm a}$Japan (Collaborating Institution), China and India (Observers) \\
} 
\end{table*}

Counting the number of published papers or 
  the number of citations of specific papers 
  which used certain telescopes is 
  one of the ways to measure the impact and importance of 
  these telescopes or facilities (e.g., \citet{davoust87, leverington96, 
  schulman97, abt98, abt00, trimble08, crabtree08, trimble09, crabtree11}),
  and analyzing citations is a method to measure the amount of impact
  of a certain paper \citep{stanek08}.

From analysis of 11,831 papers published in 20 journals of astronomy 
and astrophysics from 2001 to 2003, 
\citet{trimble08} suggested that the {\it Hubble Space Telescope (HST)}
is responsible for the largest number of optical papers,
while the most frequently cited optical papers come from the Sloan
Digital Sky Survey (SDSS), Keck, and the Anglo-Australian Telescope (AAT).
\citet{grothkopf05} also showed that the {\it HST} surpasses both 
VLT and Keck in the total number of papers, as well as in 
the numbers of papers per year (their figure 1; see also 
\citet{ringwald03, meylan04, apai10}, but see also \citet{leverington97b}).

By analysing papers resulting from optical telescopes larger than 2 m diameter
  published in 1990-1991 and cited in 1993, \citet{trimble95} and \citet{trimble96}
  found that the largest numbers of papers and citations came from
  4 m class telescopes, the Canada-France-Hawaii Telescope (CFHT)
  and AAT, followed by the Cerro-Tololo Inter-American Observatory (CTIO) 4 m,
  while the largest impact factors (five or more citations per paper per year)
  came from the University of Hawaii's 2.2 m and the Multi-Mirror Telescope 
  in Arizona \citep{trimble08}.
The analysis of papers published in 2001 and 2002,
which is after the completions of 8 m class telescopes,
\citet{trimble07} showed that the largest optical telescopes
are responsible for the largest numbers of papers,
while 4 m class telescopes displayed continued fading, 
except for the infrared United Kingdom InfraRed Telescope (UKIRT) and
InfraRed Telescope Facility (IRTF).

From the analysis of 1000 most highly-cited papers 
  published between 1991 and 1998 (125 from each year)
  and 452 astronomy papers published in {\it Nature} during $1989-1998$,
  \citet{benn01} showed that 
  the bigger the telescope, the more the paper cited,
  with citation fraction $\propto {\rm diameter}^2$.
\citet{trimble05} also suggested that big telescopes produce
more papers and more citations per paper than small ones,
from the analysis of 2100 papers produced in 2001.
\citet{ahn08} suggested that the amount of papers produced by 
  a large ($D \sim 3.6 - 10$ m) telescope is roughly 
  proportional to the diameter of its primary mirror
  (see also \citet{leverington97a}).
They also estimated the numbers of refereed and
  {\it Nature/Science} papers that might be produced by
  the Giant Magellan Telescope (GMT) annually to be 330 and 17, respectively: 
  the former by using the rough equation of $N/D \sim 14 $ 
  ($N$ is number of refereed papers
  and $D$ is the diameter of a telescope in meter)
  and the filled aperture 21.4 m of the GMT, and 
  the latter by using another rough equation of 
  $<n/A> \sim 0.05$ ($n$ is number of {\it Nature/Science} papers
  published by Keck I and each VLT telescope, and 
  $A$ is the collecting area of the primary mirror).
\citet{frogel10} initiated a series of papers to investigate what effects
  the new facilities, data archives, and means of information change
  had on astronomical publications, first by analysing the 100 most
  cited papers in each year from 2000 to 2009.

In this paper, we present an analysis of the publications
  based on results from the largest ground-based optical telescopes
  of Keck, VLT, Gemini, Subaru, and HET telescopes
  during the years of $2000-2009$.
Using the data, we try to find (i) the temporal trend
  of the publications from the above telescopes, and
  (ii) if there is any correlation in the refereed paper publications
  and the {\it Nature/Science} paper publications,
  where the latter is assumed to be the paragon of high-impact journals.
The paper is organized as follows: Sect. \ref{sect:data} describes
  the data utilized in this work.
Sect. \ref{sect:results} presents the analysis results :
  Sect. \ref{sect:number} focuses on the total number of papers and
  Sect. \ref{sect:natsci} focuses on the {\it Nature} and {\it Science}
  papers.
Finally, Sect. \ref{sect:sum} summarizes and discusses the results.

\section{Data}
\label{sect:data}

Among the telescopes with diameter larger than 8 m in Table \ref{tab1}, 
  we selected Keck, VLT, Gemini, Subaru, and HET telescopes
  for the analysis of telescope productivity,
  because these telescopes might be considered as general purpose telescopes and/or
  are well after the completion and actively produce scientific papers.
GTC, SALT, and LBT were completed after 2005,
  implying that they are still in the process of being shaken down 
  (\citet{trimble09}, their table 9).
Being completed in 1999 and producing many papers,
  HET is also included, although
  its structure is not a usual one:
  it sits at a fixed elevation angle of $55\arcdeg$ and 
  rotates in azimuth to access 81\% of the sky visible from McDonald 
  Observatory\footnote{http://www.as.utexas.edu/mcdonald/het/het\_gen\_01.html}.
Having similar structure to that of HET,
  SALT\footnote{http://www.salt.ac.za/telescope/overview/}
  also has many papers
  published\footnote{http://www.salt.ac.za/science/publications/science/}
  since its completion of 2005.

The data on the papers published by the Keck telescopes
    were obtained from the online site of 
    http:// \\ www2.keck.hawaii.edu/library/keck\_papers.html,
  those from the VLT telescopes are from
    http://archive.eso. \\ org/wdb/wdb/library/publications/form,
  those from the Gemini telescopes are from
    http://www.gemini.edu/ \\ science/publications/, 
  those from the Subaru telescope are from
    http://subarutelescope.org/Observing/ \\ Proposals/Publish/index.html, and
  those from the HET are from
    http://www.as.utexas.edu/mcdonald/het/ \\ sci\_pub.html.
We consider only ``refereed'' papers in this study,
  and we exclude any of the symposium proceedings.
For the Gemini Observatory papers, 
  the observatory webpage provides two separate pages:
  (i) `papers by users' which are based on data taken with the Gemini
  telescopes or from the Gemini Science Archive, and (ii) 
  `papers by Gemini staff' which are science and engineering papers
  published by the staff in journals and conference proceedings.
From the two sources we collected all the refereed papers 
  which have used the Gemini Observatory data for their researches,
  excluding any overlap papers.

The five databases obtained from these five observatories are then
  merged together and scrutinized to find any overlapped papers.
All these overlap papers appearing in more than one database are
  carefully examined in the full texts.
If these papers are actually produced based on the data obtained at multiple
  telescopes, then the information is kept, while it is discarded if not.
The detailed cases where two (or more) papers are kept in the final list are :
  (1) when two telescopes appear in the title (e.g. \citet{venn01});
  (2) when two telescopes appear in the footnote attached to the title
      (e.g. \citet{zheng00});
  (3) when one telescope appears in the footnote attached to the title and
      another telescope in the footnote attached to author(s) as like an affiliation
      (e.g. \citet{hu02});
  (4) when one telescope appears in the footnote attached to the title and
      another telescope in the section describing the observations
      (usually Section 2) (e.g. \citet{vreeswijk04});
  (5) when one telescope appears in the footnote attached to the title and
      a public use data and/or any existing data from another telescope is used
      (e.g. \citet{schaye00}, \citet{dessaugeszavadsky02});
  (6) when two telescopes appear in footnotes attached to authors 
      (e.g. \citet{drory01});
  (7) when two telescopes appear in the Abstract (e.g. \citet{darocha02}); and
  (8) when two telescopes appear in the main text, in the section describing 
      the observations (e.g. \citet{deBreuck01}).

There are several cases which are excluded from the final list:
  (1) one telescope is kept in the final list
      when data of the telescope are used in the analysis (e.g. Figure),
      while another telescope is not kept if only a previous study that used the telescope 
      is cited (e.g. \citet{pettini01, barth03});
  (2) when a paper used data from a telescope, while another telescope is just
      mentioned because a large program aims to get data in the future 
      with all of these telescopes, only the former is kept (e.g. \citet{fischer05});
  (3) when a different telescope in an observatory is actually used 
      instead of the large ($D > 8$ m) telescope, it is excluded from the list
      (e.g. \citet{hoeflich04}).
In spite of the careful inspection of each of the overlap papers,
  a small fraction of papers (typically $\simlt 1$\%) remain ambiguous
  if the telescope listed has actually contributed to the paper.

Since it takes long for optical telescopes today
  to ramp up to normal operations \citep{trimble09},
  it is worthwhile to check the operation start years 
  of the selected telescopes.
The two Keck telescopes are built in 1993 May and
  1996 October\footnote{http://keckobservatory.org/about/the\_observatory}.
The first light for the VLT unit 1 telescope (`Antu') was obtained 
  in late May 1998, and it went into routine scientific operation
  on 1999 April 1\footnote{http://www.eso.org/public/teles-instr/vlt.html}.
The first lights for the units 2, 3, and 4 of the VLT telescopes 
  (named `Kueyen', `Melipal', and `Yepun', respectively) were obtained
  1999 March 1\footnote{http://www.eso.org/public/news/eso9921/},
  2000 January 26\footnote{http://www.eso.org/public/news/eso0004/}, and 
  2000 September 3\footnote{http://www.eso.org/public/news/eso0028/},
  respectively.
Gemini North saw first light in 1999,
  and began scientific operations 
  in 2000\footnote{http://en.wikipedia.org/wiki/Gemini\_Observatory},
  while Gemini South opened a year later than its twin
  in 2000\footnote{http://astro-canada.ca/\_en/a2113.html}.
Subaru telescope saw first light 
  in 1999 January 28\footnote{http://www.spacetoday.org/Japan/Japan/Astronomy.html}.
Since the start of the Keck telescopes was 1993,
  it could be reasonably deduced that 
  the Keck telescopes were already in the process of normal operations
  and paper productions in 2001
  as seen in the table 9 of \citet{trimble09},
  while naturally more than one telescope brought about synergies.
On the other hand, VLT, Gemini, Subaru, and HET telescopes, built in 
  $1998-2000$, $1999-2000$, 1999, and 1999, respectively, should be still in their
  early phases in 2000 and 2001, which are confirmed 
  in Figure \ref{fig2} in the next Section.

Table \ref{tab3} shows the final paper productivities of
  the Keck, VLT, Gemini, Subaru, and HET observatories 
  during the period of $2000-2009$.
The number fraction of the excluded papers
  to the total number of papers provided on the Web by each observatory is
  typically $\simlt 1$\%.

\begin{table*} 
\begin{center}
\caption{Number of Papers Published by the Current Largest (D $>8$ m) Ground-based Optical
Telescopes During $2000-2009$}\label{tab3}
\begin{tabular}{lrrrrrrrrrrr}
\hline\hline
{Observatory$^{\rm a}$} & {2000} & {2001} & {2002}& {2003}
 & {2004} & {2005} & {2006} & {2007} & {2008} & {2009} & {Sum}\\
\hline
Keck        &161&161&176&199&208&217&265&308&262&269&2226\\
VLT         & 50&105&158&253&326&344&398&470&466&436&3006\\
Gemini      & 16& 35& 51& 50& 71& 99&136&166&162&185& 971\\
Subaru      & 17& 23& 50& 50& 68& 63& 81& 95& 88&107& 642\\
HET         &  9& 13& 10& 19& 19&  7& 19& 18& 12&  5& 131\\
\hline
Sum & 253 & 337 &445 &571 &692 &730 &899 &1057 &990 &1002 & 6976\\
\hline
\end{tabular}
\medskip\\
\end{center}
$^{\rm a}$Includes all the component telescopes : two for Keck and Gemini, four for VLT
\end{table*}

\section{Results}
\label{sect:results}

\subsection{Total Number of Papers}
\label{sect:number}

\begin{figure*}
\begin{center}
\includegraphics[width=17cm]{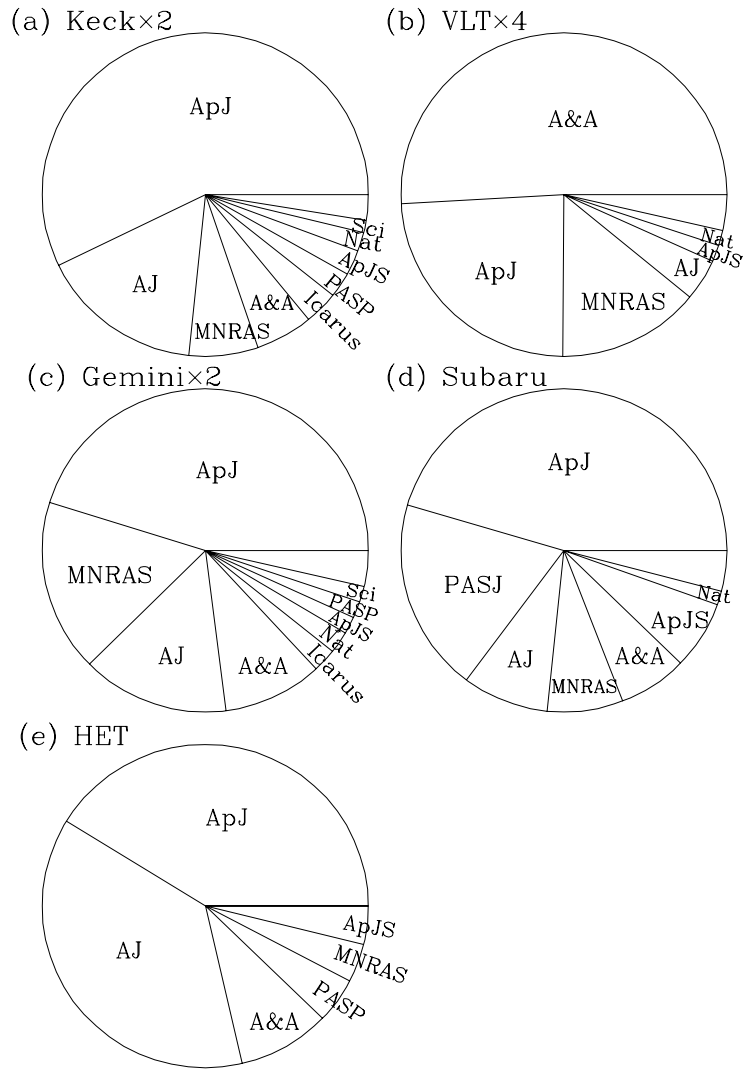} 
\caption{The distribution of journals published
  by using (a) the two Keck telescopes, (b) the four VLT telescopes,
  (c) the two Gemini telescopes, (d) the Subaru telescope, and
  (e) the HET
  during $2000-2009$ are shown in pie chart.
Journals with percentage of papers larger than 1\% are labeled, and
  the empty slots without labels in panels (a) to (d) are all other journals
  with percentages less than 1\%.
}
\label{fig1}
\end{center}
\end{figure*}

Figure \ref{fig1} shows the pie charts for the papers produced by using 
  the Keck, VLT, Gemini, Subaru, and HET telescopes,
  displaying the percentages of various journals,
  where the journals with percentage larger than 1\% are labeled.
While {\it Astrophysical Journal Supplements} ($ApJS$) is shown separately, 
  {\it Astrophysical Journal} ($ApJ$) include 
  the {\it Astrophysical Journal Letters} ($ApJL$) publications
  (see \citet{frogel10} for the history on the separation of $ApJ$ and $ApJL$).
Keck telescopes published 57.2\% of papers in $ApJ$ (including $ApJL$)
  (it becomes 59.7\% if $ApJS$ is also included),
  and 75.9\% in the American journals of $ApJ$, $ApJS$, and {\it Astronomical Journal} ($AJ$).
While Gemini telescopes published 45.2\% of papers in $ApJ$,
  the value becomes 59.8\% if $AJ$ is also included.
Subaru telescope has 60.9\% of papers published in
  the American journals of $ApJ$, $ApJS$, and $AJ$, while
  19.3\% of papers are published in the
   {\it Publications of the Astronomical Society of Japan} ($PASJ$).
HET published 78.6\% of papers in $ApJ$ and $AJ$, 
  and 82.4\% in the American journals of $ApJ$, $AJ$, and $ApJS$.
The ESO VLT observatory, however, has published dominantly 
  in the European journal {\it Astronomy and Astrophysics} ($A\&A$) (50.9\%) and
  the UK journal {\it Monthly Notices of the Royal Astronomical Society} ($MNRAS$) (14.2\%),
  while still 29.8\% of papers are published in the American journals 
  of $ApJ$, $AJ$, and $ApJS$ \citep{abt10}.
While 12.5\% of Keck telescope papers are published in $A\&A$ and $MNRAS$,
  29.8\%                         of VLT papers are published in 
  the American journals.
While \citet{abt10} notes that most (55\%) of the astronomical articles
  in journals with impact factors \citep{frogel10} greater than 2.0 are 
  published in just four journals of $A\&A$, $AJ$, $ApJ$ 
  (including $ApJL$ and $ApJS$), and $MNRAS$,
  the percentages of papers published in these journals
  by using the Keck, VLT, Gemini, Subaru, and HET telescopes are
  88.5\%, 94.9\%, 88.7\%, 75.2\%, and 95.4\%, respectively.
If $PASJ$ is also included, the percentage for the Subaru increases to
  94.5\%.

\begin{figure*}
\begin{center}
\includegraphics[width=17cm]{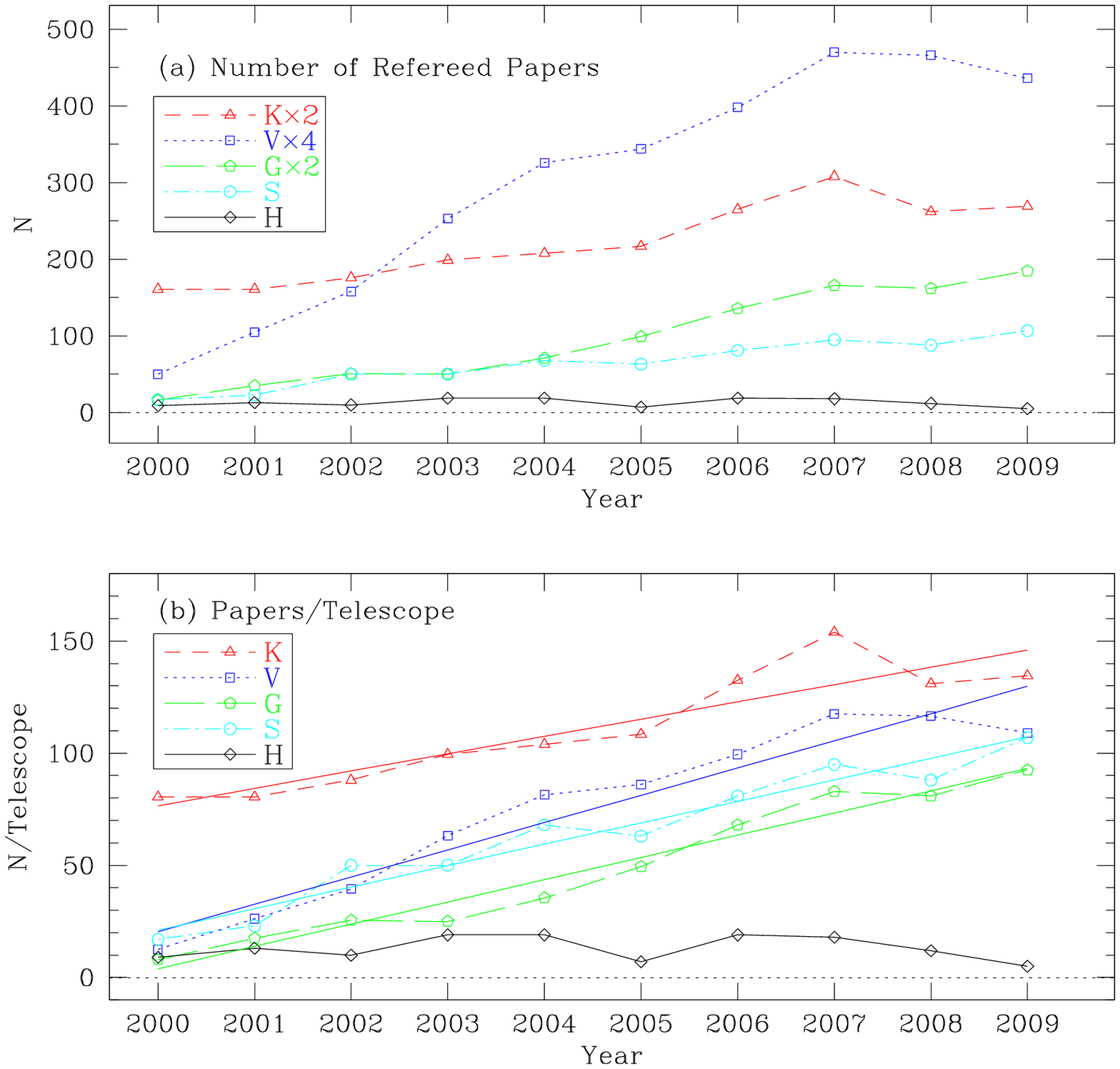}
\caption{(a) Yearly distribution of the refereed papers published
by using the Keck, VLT, Gemini, and Subaru telescopes
for the period of $2000-2009$.
`K$\times2$' (triangles) is for the two Keck telescopes,
`V$\times4$' (squares) for the four VLT telescopes,
`G$\times2$' (pentagons) for the two Gemini telescopes,
`S' (circles) is for the Subaru telescope, and
`H' (diamonds) is for the Hobby-Eberly Telescope.
(b) The numbers of papers in panel (a) are divided by
2, 4, and 2 for Keck, VLT, and Gemini telescopes, respectively,
to show the number of papers produced by each telescope.
Non-weighted least squares fits to the data
  are shown as solid lines, and
  the slopes for the Keck, VLT, Gemini, and Subaru telescopes
  are $7.8 \pm 1.2$, $13.1 \pm 1.2$, $12.6 \pm 1.0$, and
  $9.6 \pm 0.8$, respectively.
}
\label{fig2}
\end{center}
\end{figure*}

Figure \ref{fig2} (a) shows the yearly distribution of the refereed papers 
  produced by using the Keck (triangles), VLT (squares), Gemini (pentagons), 
  Subaru (circles), and HET (diamonds) observatories 
  for the period of $2000-2009$.
The two Keck telescopes (`K$\times2$', triangles) show very stable,
  but still ascending slope in the number of papers up to 2007
  (N(2007)$=308$),
  after which it slightly decreased.
The four VLT telescopes (`V$\times4$', squares) show rapid increase 
  in the number of papers from 2000 to 2007 and again
  the value of 2007 (N$=470$) is the maximum during the ten years.
In 2003, the number of papers produced by using VLT 
  crossed over that by using Keck \citep{grothkopf05, trimble08}.
Built in 1999 and 2000, the two Gemini telescopes also show
  rapid increase in the number of paper from 2000 (N$=16$) 
  to 2009 (N$=185$), while the latter is the maximum number
  among the ten years.
Having only one 8.2 m telescope unlike others above,
  the Subaru telescope shows steady increase 
  in the number of papers, and the maximum value is
  in 2009 (N$=107$).
Unlike the above telescopes, the 9.2 m HET
  shows almost steady value of $13.1 \pm 5.4$ for the number of papers
  each year from 2000 to 2009.
While \citet{abt10} showed that the astronomical research rates
  in the US, the UK, and Europe have not reached a maximum and
  seem still increasing,
  it will be needed to gather data for at least a few more years in the future
  to see if it is the same for the publications from the above telescopes
  since some telescopes show leveling off or even decrease in 
  the number of papers after 2007.
It is worth here to note that \citet{frogel10}(his figure 1) found 
  the rise of the total number of authors for the top 100 papers
  during the period of $2000-2009$ is steep from 2000 to 2007
  and levels off from 2007 to 2009.
As he writes that it is not easy to determine if such a rise is 
  typical for all astronomical articles or is confined to 
  the top 100 for each year, it could be interested to find
  if there is any correlation in the leveling off of
  both the total number of papers from the largest optical telescopes
  and the total number of authors for the top 100 papers
  after 2007.

Since the Keck, VLT, and Gemini observatories have 
  two, four, and two telescopes, respectively,
  it is not fair to compare with each other and 
  also with the Subaru and the HET.
We, therefore, divided the total number of papers produced by 
  the Keck, VLT, and Gemini observatories by 2, 4, and 2, respectively,
  and showed the result in Figure \ref{fig2} (b),
  which shows the number of papers produced by using each individual telescope.
Unlike the other telescopes, each telescope of the Keck observatory (`K') 
  shows the largest values in the numbers of papers. 
After 2002, VLT keeps the top position 
  among the three 8 m class telescopes of VLT, Subaru, and Gemini.
While each data point is connected by broken lines,
  we showed the non-weighted least squares fit results
  by solid lines and obtained the slopes for each of the
  Keck, VLT, Gemini, and Subaru telescope as
  $7.7 \pm 1.2$, $12.1 \pm 1.2$, $9.9 \pm 0.7$, and
  $9.6 \pm 0.8$, respectively.
VLT shows the largest value in the slope, and
  Gemini, Subaru, and Keck follow.
VLT could have the largest slope, probably because it has (i)
  VLTI (VLT interferometer) : the capability of combining all the
  telescopes also using smaller auxiliary telescopes
  ($\sim 4$ \% of VLT papers are from the VLTI),
  (ii) powerful suite of instruments,
  (iii) data reduction pipelines,
  (iv) a queue-based observing system (about half the time),
  (v) data archive,
  (vi) synergy of having largest number of same telescopes at the same place
  and/or 
  (vii) good (especially technical) support \citep{grothkopf05}.
The reality is that it is probably some combination of these possibilities,
  and of course the common factor is ESO's larger operations budget.
The fact that each of the 10 m Keck telescopes produce a larger number of
  papers than any each of the other 8 m class telescopes is consistent with
  the finding of \citet{ahn08} that $N \propto D$ for large optical telescopes
  (where, $N$ is number of refereed papers
  and $D$ is the diameter of a telescope in meter)
  (see also \citet{abt80,leverington97a}).
This, on the other hand, could result from the fact that
  almost every aspect of a telescope project is scaling 
  with the telescope diameter: its construction budget, its operational budget,
  the user community, the level of user support, etc.

The order of the slope values of the fittings of the number of refereed papers
  over year for each of the telescopes could be explained by
  other parameters, e.g., the number of instruments of the telescopes.
Currently, the number of instruments of the five telescopes are :
  VLT - 12 (FORS1, FORS2, ISAAC, UVES, NCAO, VIMOS, FLAMES, VISIR, SINFONI,
    CRIRES, HAWK-I, and X-shooter),
  Gemini - 11 (Altair, GMOS, GNIRS, Michelle, NIFS, NIRI,
    FLAMINGOS-2, GMOS, NICI, Phoenix, and T-ReCS),
  Keck - 9 (HIRES, LRIS, NIRC,
    DEIMOS, ESI, NIRC2, NIRSPEC, NIRSPAO, and OSIRIS),
  Subaru - 8 (AO188, COMICS, FMOS, FOCAS, HDS, IRCS, MOIRCS, and Suprime-Cam),
  and HET - 3 (LRS, MRS, and HRS).
This decreasing order of the number of instruments from VLT (12) to HET (3)
  is almost similar to that of the slopes above, i.e.
  VLT ($12.1 \pm 1.2$),
  Gemini ($9.9 \pm 0.7$),
  Subaru ($9.6 \pm 0.8$),
  Keck ($7.7 \pm 1.2$), and almost flat HET.
The existence of data archives for the observatories of VLT (http://archive.eso.org/),
  Gemini (http://www3.cadc-ccda.hia-iha.nrc-cnrc.gc.ca/gsa/), and
  Subaru (http://smoka.nao.ac.jp/) coincide with the highest values
  for the slopes of these three telescopes, especially for the Subaru
  having larger slope than does Keck.

\begin{figure*}
\begin{center}
\includegraphics[width=17cm]{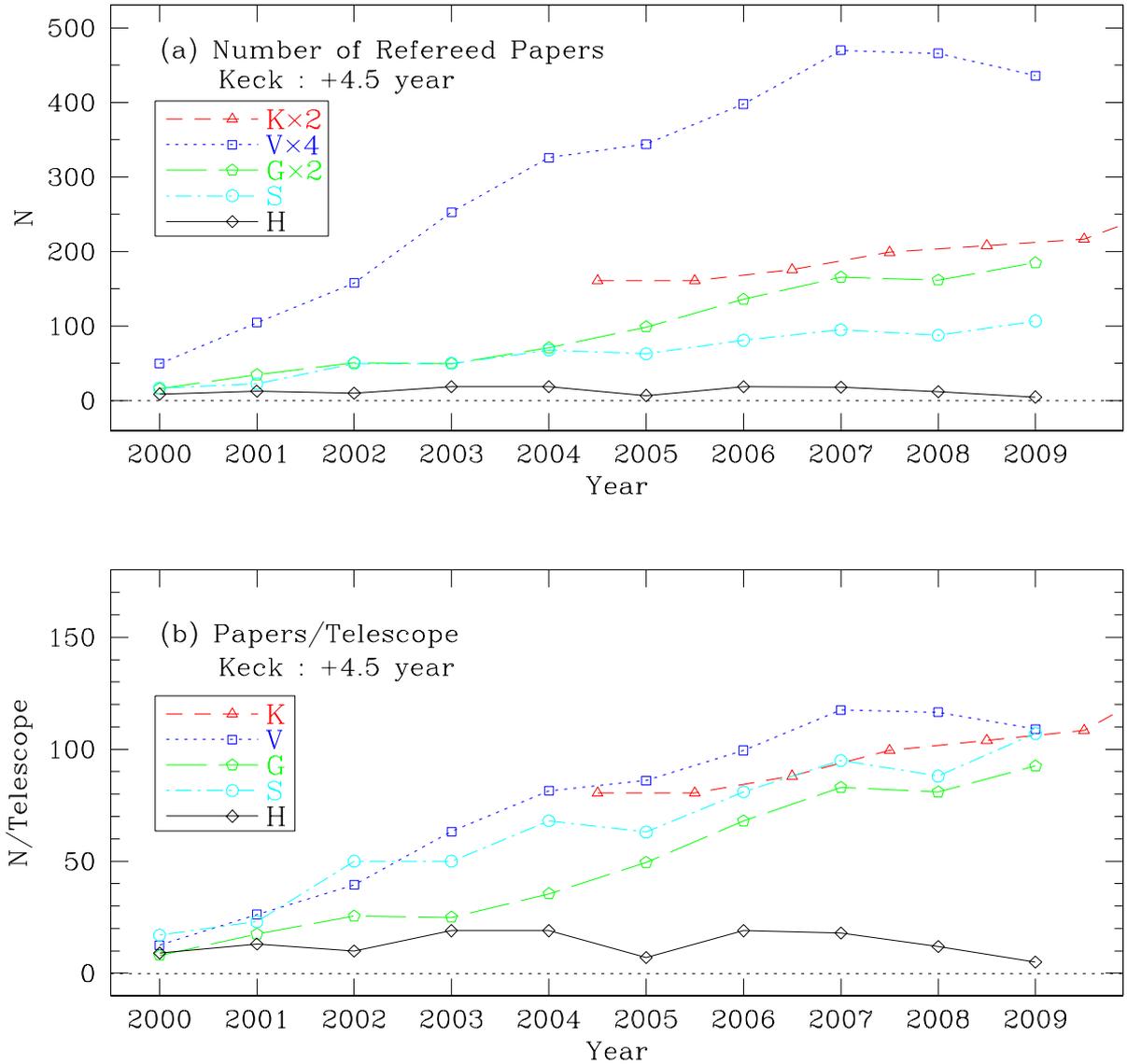}
\caption{Same as in Figure \ref{fig2}, but the data 
  for Keck are shifted toward +4.5 years, so that
  all telescopes approximately have almost similar 
  zero point in time, i.e., beginning of the science operations.
}
\label{fig3}
\end{center}
\end{figure*}

The fact, however, that Keck shows the largest values in the number of
  refereed papers per unit telescope could also be the result of
  earlier starts of the Keck telescopes than those of other telescopes, 
  so that they could be in more stable operations.
This is confirmed in Figure \ref{fig3}, where the data for Keck
  are shifted forward +4.5 years (the mean of the start years of two Keck
  telescopes of 1993 and 1996 is taken to be 1994.5)
  assuming that VLT, Gemini, Subaru, and HET started at around 1999.
Panel (b) of Figure \ref{fig3} shows that, if the telescopes of VLT, Gemini,
  and/or Subaru have similar time for them to be stable enough
  for scientific operations as in the case of Keck,
  each component of them would have almost similar productivity
  as that of each of the Keck telescopes,
  especially for the four VLT telescopes.

\begin{figure*}
\begin{center}
\includegraphics[width=17cm]{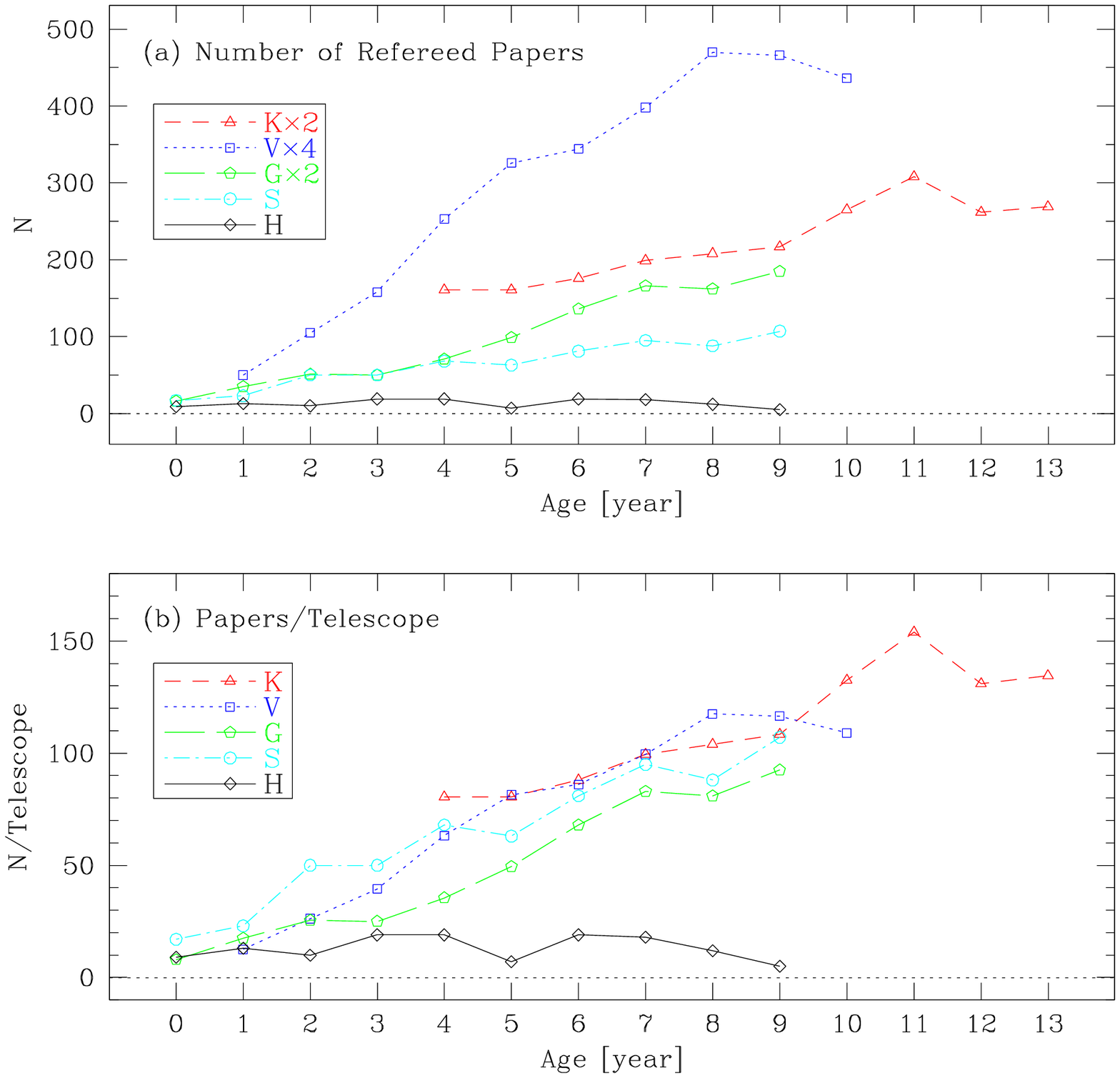}
\caption{Same as in Figure \ref{fig2}, but the horizontal axis
  is for the ages of the observatories.
The age is set to zero when the first paper using the telescope
  is published.
}
\label{fig4}
\end{center}
\end{figure*}

Another way to compare the productivities of telescopes 
  is to look at the number of papers as a function of age, 
  where the `age' is set to zero when the first paper 
  using the telescope is published
  (Keck - 1996 January; VLT - 1999 March; 
  Gemini - 2000 December; Subaru - 2000 February; HET - 2000 January).
This variation versus the different ages 
  is shown in Figure \ref{fig4}.
Figure \ref{fig4} (a) shows almost same results 
  as in Figure \ref{fig3} (a).
Figure \ref{fig4} (b), however, shows good progress for each of the Keck telescopes 
  while, for the near futures of VLT, Subaru, and Gemini,
  more data is needed more data to see 
  if they will show an increase in productivity as for the ages of 
  [9, 11] of Keck or leveling off as for the ages of [11, 13] of Keck.

\subsection{Nature and Science Papers}
\label{sect:natsci}
Citations to the papers are usually considered as the typical measure of
  the impact that journals/papers bring about (e.g. \citet{apai10}) and
  \citet{frogel10} showed that {\it Nature} and {\it Science}
  are not included in the five journals ($A\&A, AJ, ApJ, ApJS$, and $MNRAS$)
  that account for 80 to 85\% of the total citations for each year.
These two journals, however, still hold the highest 
  impact factors\footnote{The impact factors for {\it Nature} and {\it Science}
  for the years of 2006, 2007, 2008, and 2009 are 
  26.681, 28.751, 31.434, and 34.480 and
  30.028, 26.372, 28.103, and 29.747, respectively,
  which is released by Journal Citation Reports, Thomson Reuters.}, and
  are generally regarded as paragon of high-impact journals 
  (see, e.g., \citet{metcalfe05}).
Here, we assume that the publications in {\it Nature} and {\it Science}
  are the prototype of high impact papers in astronomy, 
  meaning any of new discoveries, breakthroughs in a specific field of astronomy, or
  new findings for celestial objects/phenomena.
It is true, however, that there are opposite opinions
  on the journals of {\it Nature} and {\it Science}
  that they are too sensational.
In spite of the fact that these two journals are highly ranked
  by Thomson Reuters,
  the Thomson Reuters Institute for Scientific Information (ISI) uses
  citation metrics only as one indicator among others to predict 
  Nobel prizewinners.
Since `of the 28 physics Nobel prizewinners from 2000 to 2009, just 5 are
  listed in ISI's top 250 most-cited list for that field' \citep{frey10}.

\begin{figure*}
\begin{center}
\includegraphics[width=17cm]{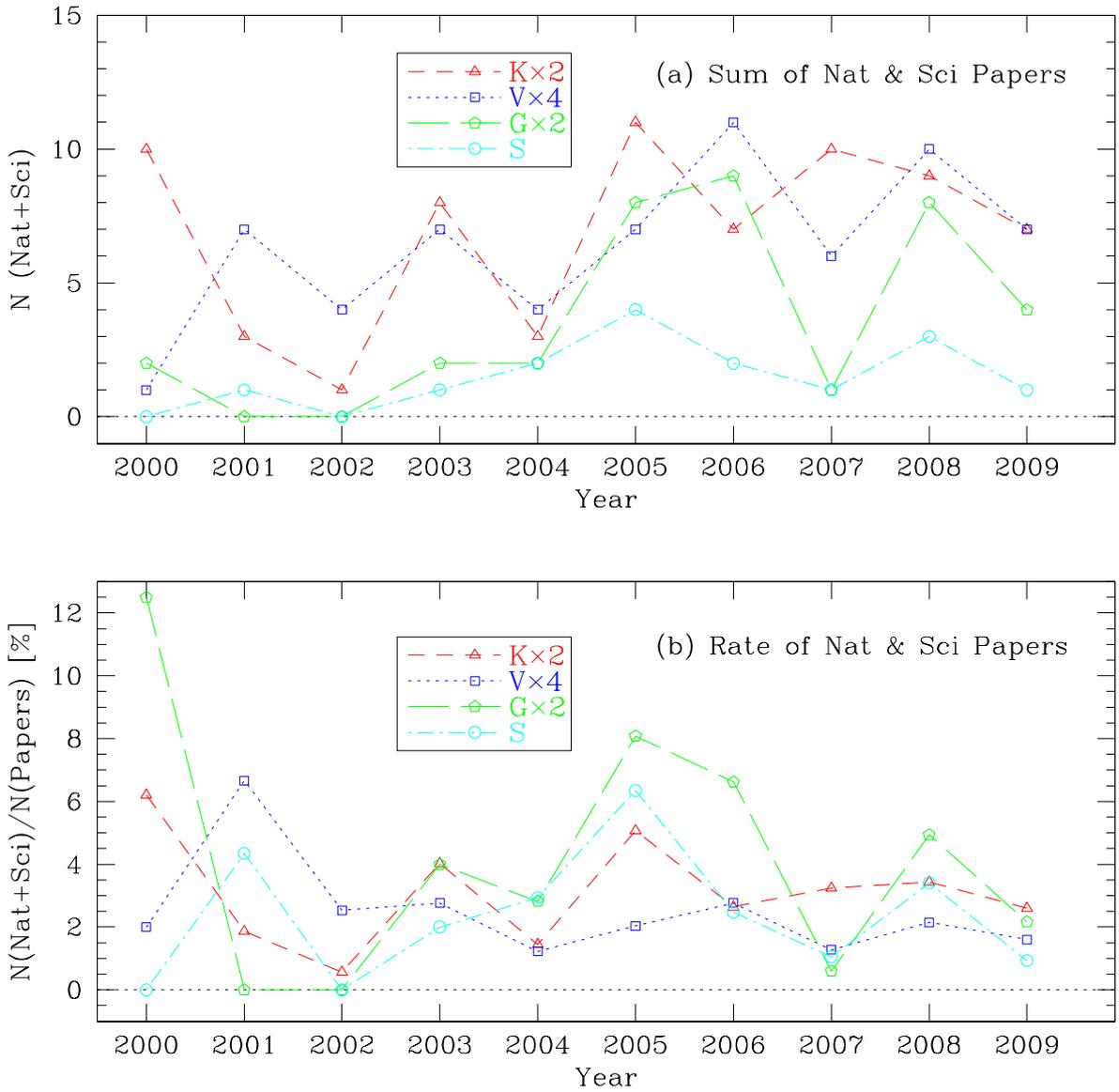}
\caption{(a) Yearly distribution of the number of papers published in
  {\it Nature} and {\it Science}
  by using the Keck, VLT, Gemini, and Subaru telescopes
  for the period of $2000-2009$.
`K$\times2$' (triangles) is for the two Keck telescopes,
`V$\times4$' (squares) for the four VLT telescopes,
`G$\times2$' (pentagons) for the two Gemini telescopes, and
`S' (circles) is for the Subaru telescope.
(b) The rate of the number of papers published
  in {\it Nature} and {\it Science}
  divided by the total number of refereed papers
  for the Keck, VLT, Gemini, and Subaru telescopes
  for the period of $2000-2009$.
Symbols are same as in panel (a).
}
\label{fig5}
\end{center}
\end{figure*}

Figure \ref{fig5} (a) shows the yearly distribution of the number 
  of papers published in {\it Nature} and {\it Science}
  by using the Keck, VLT, Gemini, and Subaru telescopes
  for the period of $2000-2009$, and 
  Figure \ref{fig5} (b) shows the rate of 
  {\it Nature} and {\it Science} papers among
  all the refereed papers produced by using the telescopes.
Table \ref{tab4} shows the statistics of these papers,
  where the upper part is for the number of papers and
  the lower part is for the rate of {\it Nature} and {\it Science} papers
  among all the refereed papers produced by using the telescopes.
The Keck telescopes produce the largest mean (N$=6.9 \pm 1.1$, $\sigma=3.4$) 
  number of {\it Nature} and {\it Science} papers, and
  the VLT (mean N$=6.4 \pm 0.9$, $\sigma = 2.9$) follows it.
Gemini shows a somewhat larger fluctuation like Keck 
  (mean N$=3.6 \pm 1.1$, sigma $=3.5$), and
  Subaru shows rather low, but still steady distribution
  (mean N$=1.5 \pm 0.4$, sigma $=1.3$).
The mean rates of {\it Nature} and {\it Science} papers
  among all the refereed papers produced by these observatories
  are between 2.4 and 4.2,
  while the median values are 2.1 (VLT), 2.5 (Subaru), 3.2 (Keck), and
  4.0 (Gemini).
Gemini still shows larger value of dispersion ($\sigma = 4.0$)
  in the rate of {\it Nature} and {\it Science} papers
  among all the refereed papers
  than the other observatories ($\sigma = 1.6, 1.7,$ and 2.0
  for VLT, Keck, and Subaru, respectively).
The reason why there is no {\it Nature} and/or {\it Science} papers
  from the HET could be attributed, among others,
  to the small number of papers based on the HET data and
  possibly to the fact that its structure is designed in a very special way
  (see Section 1), specifically for spectroscopy, at very low cost.

\begin{table} 
{\footnotesize 
\begin{center}
\caption{Statistics of {\it Nature} and {\it Science} Papers During $2000-2009$}\label{tab4}
\begin{tabular}{lccccc}
\hline\hline
{Observatory} & {Min} & {Max} & {Mean}& {$\sigma$}& {Median}\\
\hline
Keck (N)    &  1& 11& $6.9 \pm 1.1$ & 3.4 & 8 \\
VLT (N)     &  1& 11& $6.4 \pm 0.9$ & 2.9 & 7 \\
Gemini (N)  &  0&  9& $3.6 \pm 1.1$ & 3.5 & 2 \\
Subaru (N)  &  0&  4& $1.5 \pm 0.4$ & 1.3 & 1 \\
            &   &   &               &     &   \\
Keck (\%)   &  0.6&  6.2& $3.1 \pm 0.5$ & 1.7 & 3.2 \\
VLT (\%)    &  1.2&  6.7& $2.5 \pm 0.5$ & 1.6 & 2.1 \\
Gemini (\%) &  0.0& 12.5& $4.2 \pm 1.3$ & 4.0 & 4.0 \\
Subaru (\%) &  0.0&  6.3& $2.4 \pm 0.6$ & 2.0 & 2.5 \\
\hline
\end{tabular}
\end{center}
} 
\end{table}

\begin{figure*}
\begin{center}
\includegraphics[width=17cm]{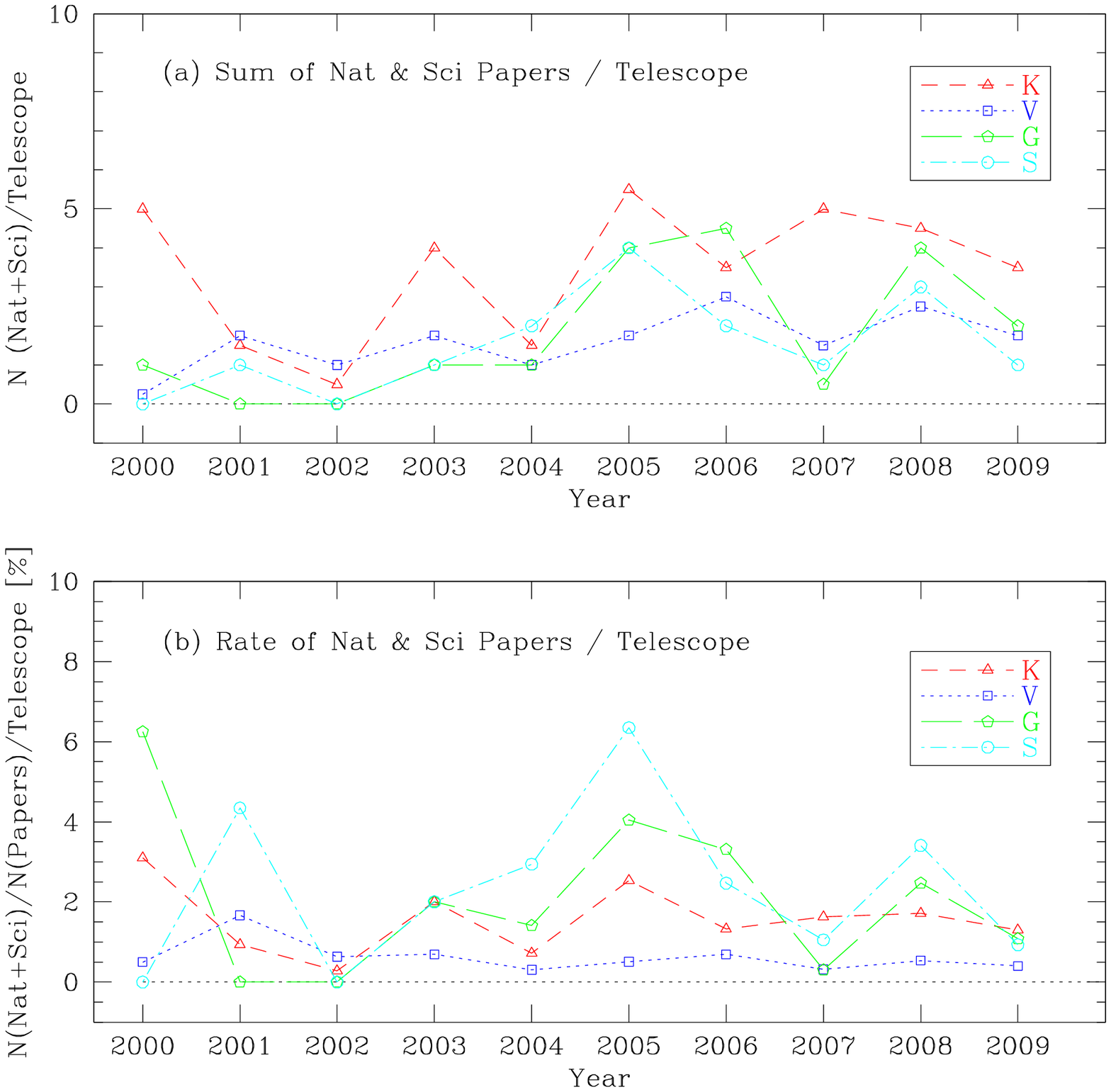}
\caption{(a) Yearly distribution of the number of papers published in
  {\it Nature} and {\it Science}
  by using each of the Keck, VLT, Gemini, and Subaru telescopes
  for the period of $2000-2009$.
`K' (triangles) is for each of the two Keck telescopes,
`G' (pentagons)    for each of the two Gemini telescopes,
`V' (squares)      for each of the four VLT telescopes, and
`S' (circles) is for the Subaru telescope itself.
(b) The rate of the number of papers published
  in {\it Nature} and {\it Science}
  divided by the total number of refereed papers
  for each of the Keck, VLT, Gemini, and Subaru telescopes
  for the period of $2000-2009$.
Symbols are same as in panel (a).
}
\label{fig6}
\end{center}
\end{figure*}

In Figure \ref{fig5} (a) the primary reason why the Subaru Observatory
  shows least number of {\it Nature} and {\it Science} papers 
  compared to the other observatories is 
  because the number of component telescopes is different.
We, therefore, plotted in Figure \ref{fig6} (a)
  the yearly distribution of the number 
  of papers published in the two journals of {\it Nature} and {\it Science}
  by using each of the Keck, VLT, Gemini, and Subaru telescopes
  for the period of $2000-2009$.
Figure \ref{fig6} (b) shows the rate of
  {\it Nature} and {\it Science} papers among
  all the refereed papers produced by using each of the telescopes.
Table \ref{tab5} shows the statistics of these papers,
  where the upper part is for the number of papers and
  the lower part is for the rate of {\it Nature} and {\it Science} papers
  among all the refereed papers produced by using each of the telescopes.

\begin{table} 
{\footnotesize 
\begin{center}
\caption{Statistics of {\it Nature} and {\it Science} Papers Produced
  by Using Each Telescope During $2000-2009$}  \label{tab5}
\begin{tabular}{lccccc}
\hline\hline
{Telescope} & {Min} & {Max} & {Mean}& {$\sigma$}& {Median}\\
\hline
Each Keck (N)    &0.5&5.5& $3.5 \pm 0.5$ & 1.7 & 4 \\
Each VLT (N)     &0.3&2.8& $1.6 \pm 0.2$ & 0.7 & 1.8 \\
Each Gemini (N)  &  0&4.5& $1.8 \pm 0.5$ & 1.7 & 1 \\
Subaru (N)       &  0&  4& $1.5 \pm 0.4$ & 1.3 & 1 \\
                 &   &   &               &     &   \\
Each Keck (\%)   & 0.3& 3.1& $1.6 \pm 0.3$ & 0.8 & 1.6 \\
Each VLT (\%)    & 0.3& 1.7& $0.6 \pm 0.1$ & 0.4 & 0.5 \\
Each Gemini (\%) & 0.0& 6.3& $2.1 \pm 0.6$ & 2.0 & 2.0 \\
Subaru (\%)      & 0.0& 6.3& $2.4 \pm 0.6$ & 2.0 & 2.5 \\
\hline
\end{tabular}
\end{center}
} 
\end{table}

Compared to the other 8 m class telescopes 
  (mean values of N$=1.5 - 2.3$, median N$=1 - 1.8$),
  the 10 m Keck telescope shows the largest
  mean (N$=3.5 \pm 0.5$, $\sigma = 1.7$) and median (N$=4$) number of papers.
While each telescope of the observatories shows similar rate of 
  {\it Nature} and {\it Science} papers
  among all the refereed papers produced by using each of the telescopes
  (mean N$=0.7 - 2.3$, median N$=0.6 - 2.5$),
  VLT shows the lowest rate and Subaru shows the highest rate.
Since the total numbers of refereed papers produced by each
  of the VLT telescopes is not small (see Figure \ref{fig2} (b)),
  it could be concluded that
  the users of the VLT telescopes tend to publish more papers
  in the usual astronomical journals than in the journals of
  {\it Nature} and {\it Science}.
From each of the Keck, VLT, Gemini and Subaru telescopes,
  the overall mean and median values in Table \ref{tab5}
  are N$=2.1 \pm 0.9$ and N$=2.0 \pm 1.4$, respectively, 
  for the number of {\it Nature} and {\it Science} papers and
  $1.7 \pm 0.8$\% and $1.6 \pm 0.9$\%, respectively,
  for the rates.
We, therefore, could conclude that each of the current 8 to 10 m class telescopes
  is producing $2.1 \pm 0.9$ {\it Nature} and {\it Science} papers annually and
  the rate of these papers among all the refereed papers
  produced by using that telescope is $1.7 \pm 0.8$\%.

In the meanwhile, it is necessary to note that these statistics represent
  only the current trend considering the number of the active, fore-front 
  astronomical facilities including the largest (D $>8$ m) ground-based optical telescopes,
  space telescopes, specially designed/special purpose telescopes, etc.
  and the policies of the {\it Nature} and {\it Science} journals
  regarding the balance among the different disciplines represented in the journals.
In the next decade, some, many, or most of the above optical/infrared telescopes 
  currently producing many {\it Nature} and {\it Science} papers
  will be probably substituted by ELTs in the sense that 
  producing new discoveries and doing highest impact sciences at that time.

Telescopes smaller than ELTs could get some ideas
  on their long-term performance from the case of CFHT.
CFHT is one of the most competitive telescopes among the 4 m class 
  \citep{benn01}, and 
  has produced around 130 refereed papers in 2010 \citep{veillet11}
  in the current era of large (D $\sim 8-10$ m) optical telescopes
  though its being only 3.6 m in diameter.
The annual number of refereed publications based significantly on CFHT 
  has been more or less over 50 since 2000, and it already became larger than 100
  in 2007\footnote{CFHT Annual Report 2007 :
  http://www.cfht.hawaii.edu/AnnualReports/AR2007e\_web.pdf}.
This productive trend of results might be based on the efforts 
  made by the Observatory like the followings : 
\begin{itemize}
  \item {\em} Queued Service Observing (QSO) mode affords as much real observing time
  as requested by the observers.  
  The QSO personnel select the observing conditions according to 
  the sky clearance and seeing so that the optimum 
  condition is given to every successful observing proposal.
  \item {\em} The obtained data are provided to the principal investigators
  after the preprocessing is finished, so that observers do not need to
  spend a minute on it.
  \item {\em} CFHT affords high-performance wide-field imagers 
  (field of view of $0.96\arcdeg \times 0.94\arcdeg$ in optical, and
  $20\arcmin \times 20\arcmin$ in near-infrared wavebands), 
  which is a big advantage of this telescope 
  making it possible for this telescope to achieve high level of paper production
  \citep{cuby07}.
  \item {\em} CFHT carries out large programs
  and collaborative observing projects with many other facilities,
  which maximized the value of the telescope.
\end{itemize}

\section{Summary and Discussion}
\label{sect:sum}

We have analysed the ten year ($2000-2009$) publication record of the current
  largest (D $>8$ m) ground-based optical telescopes of Keck, VLT, Gemini, Subaru, and HET.
During the ten year period, the telescopes of Keck, VLT, Gemini,
  and Subaru showed increasing numbers of refereed papers and this tendency is
  still preserved when we divided the number of papers
  by the number of telescope components (2 for Keck and Gemini, and
  4 for the VLT telescopes).
Each telescope of the Keck, VLT, Gemini, Subaru, and HET observatories
  produced 135, 109, 93, 107, and 5 refereed papers, respectively,
  in 2009.
For the ten year period,
  the number of papers produced by each of the telescopes is largest
  for the Keck, while the largest slope in the change of
  the annual number of papers is for the VLT.
It is worthwhile to note that the impact of papers based on archival data
  can have a significant impact on a telescope's productivity. 
For example, almost half of papers published using $HST$ data are
  based on at least some archival
  data\footnote{http://archive.stsci.edu/hst/bibliography/pubstat.html},
  and this could be also a factor for VLT, Gemini, and Subaru as mentioned in \S 3.1.
While the astronomical literature continues to grow exponentially 
  by $2-3$\% \citep{frogel10},
  4\% \citep{abt98}, 
  5\% \citep{white07,trimble08}, $6-7$\% \citep{abt10}, or
  8.8\% \citep{abt95} annually,
  we will need more data for the next several years to see
  if the number of papers produced by using the telescopes
  will still increase or not,
  since some of the telescopes (Keck and VLT) show somewhat less publications
  in 2008 and 2009 than the year of 2007 (see Figure \ref{fig2} (a)).

For the papers published in the two multi-disciplinary, high-impact journals 
  of {\it Nature} and {\it Science} \citep{frogel10},
  we have shown that each telescope of the Keck, VLT, Gemini, and Subaru observatories
  is producing $2.1 \pm 0.9$ {\it Nature} and {\it Science} papers annually and
  the rate of these papers among all the refereed papers
  produced by using that telescope is $1.7 \pm 0.8$ \%.
Extending this relation obtained from the current largest
  ground-based optical telescopes,
  we may be able to conclude that this ratio of the number of
  {\it Nature} and {\it Science} papers over the number of whole
  refereed papers that will be produced by future ELTs
  of GMT, TMT and E-ELT will remain similar.
If, therefore, one of the future larger telescopes produces,
  e.g., 330 refereed papers annually,
  the above $simple$ calculation suggests that $\sim 6$
  {\it Nature} and/or {\it Science} papers might be included 
  in these publications annually.

From the comparison of the publication trends of the telescopes,
  we may conclude the followings :
\begin{itemize}
  \item {\em} While the telescope productivity means papers per telescope,
  it is expected that the more telescopes of the same kind 
  at the same location, 
  the more synergies occur.
  This includes the effectiveness of maintenance, 
  less number of observatory personnel, less cost for the facilities
  and more chances to use the instruments that are made to be attached to the same telescope.
Although this fact might not be the critical factor 
  for telescope productivity,
  the specific example of the VLT is worth noting.
The four VLT telescopes currently have the largest number of 
  instruments (12 ; see \S 3.1, four of the instruments can be used
  at the same time), largest number of papers among the telescopes considered in this study,
  and largest slope value ($12.1 \pm 1.2$) of the fitting of the number of refereed papers
  over year.

  \item {\em} The important factors that influence the growth rate of
  paper production are ramp-up in efficient operations,
  reliable instruments, useful instruments, and
  the number of good instruments available at the telescope.
  The latter point might be supported by the fact that
  the order of the number of instruments
  is almost the same as that of the slope values
  of the fitting of the number of refereed papers over year :
  VLT    (12, $12.1 \pm 1.2$),
  Gemini (11, $9.9 \pm 0.7$),
  Keck   (9, $7.7 \pm 1.2$),
  Subaru (8, $9.6 \pm 0.8$), and
  HET    (3, almost flat), where (number of instruments, the slope value).
\end{itemize}

Although it might not be able to have more than one telescope at a site
  to maximize the productivity,
  it is a natural and necessary way to increase the number of good instruments available
  and to afford the archive to maximize the use of the data.
There are also many other items that affect the productivities of telescopes, such as :
\begin{itemize}
  \item {\em} user base of the telescope,
  \item {\em} publication traditions of journals (e.g., US vs European journals) 
    \citep{schulman97, abt98, abt10, frogel10}, and
  \item {\em} support on the telescope users (pool of the management personnel),
\end{itemize}
  of which investigation in the future might give us more lessons,
  although it is not easy to get some of the data (e.g. budget).
The first item above, the user base of the telescope, might be correlated with
  the telescope subscription rate.
\citet{frogel10} showed that (in his figure 1) the membership numbers for
  the American Astronomical Society (AAS) have stayed flat for the last $10-20$ years
  (cf. \citet{abt00}),
  while those for the International Astronomical Union increased about 20\%
  over the same period.
This almost constant number of AAS membership shows no correlation
  with the increase of the numbers of papers of US telescopes of Keck, 
  HET (90\% portion for US) and Gemini (48\% portion for US) as shown in this study.
This indicates that the analysis of the user base of the large optical telescopes
  might need the database of actual telescope users,
  needless to say that of optical astronomers of the countries
  that operate the telescopes, specifically for VLT, Gemini, and HET 
  which are being operated by two or more countries.
The rather detailed analysis on the journal of AJ by \citet{bracher99} might represent 
  the studies on the publication traditions of journals.

\section*{Acknowledgments}
S.C.K. would like to thank the referee, Dennis Crabtree, for 
  providing prompt and thoughtful comments and
  detailed comments on English that greatly helped improve
  the original manuscript.
S.C.K. is also grateful to Dr. Hong Soo Park for his kind help in using
  the SuperMongo and helpful discussion,
  and to Drs. Jaemann Kyeong and A-Ran Lyo
  for their kind encouragements.
S.C.K. is a member of the Dedicated Researchers
  for Extragalactic AstronoMy (DREAM) team in Korea Astronomy and Space 
  Science Institute (KASI).

\section{Appendix}

The following five tables are squeezed into one table (Table 3) in the PASA edition,
which are shown here in arXiv version only.

\begin{table*} 
\begin{center}
\caption{Number of Papers Published by Using the Keck Telescopes During $2000-2009$}\label{tab4}
\begin{tabular}{lrrrrrrrrrrr}
\hline\hline
{Journal} & {2000} & {2001} & {2002}& {2003}
 & {2004} & {2005} & {2006} & {2007} & {2008} & {2009} & {Sum}\\
\hline
ApJ         & 83& 81& 97&109&114&131&157&183&152&166&1273\\
AJ          & 36& 42& 44& 37& 40& 30& 34& 33& 34& 31& 361\\
MNRAS       &  5&  8&  6& 15& 11& 11& 19& 34& 26& 17& 152\\
A\&A        & 11$^{\rm a}$& 17&  9&  4& 12&  7& 17& 16& 13& 21& 127\\
Icarus      &  1&  2&  6&  6&  6& 11& 10& 14& 11&  8&  75\\
PASP        &  7&  3&  9&  5&  2&  3& 10&  6&  3&  8&  56\\
ApJS        &  4&  2&  4&  6& 10&  7&  3&  8&  6&  6&  56\\
Nature      &  7&  1&  1&  6&  2&  6&  4&  4&  5&  6&  42\\
Science     &  3&  2&  0&  2&  1&  5&  3&  6&  4&  1&  27\\
NewAR       &  0&  2&  0&  2&  0&  1&  4&  0&  2&  0&  11\\
Ap\&SS      &  0&  1&  0&  3&  0&  0&  0&  1&  0&  2&   7\\
JGR         &  2&  0&  0&  2&  1&  0&  1&  1&  0&  0&   7\\
ARA\&A      &  0&  0&  0&  0&  0&  2&  2&  0&  0&  1&   5\\
GeoRL       &  0&  0&  0&  0&  3&  1&  0&  1&  0&  0&   5\\
AdSpR       &  0&  0&  0&  1&  1&  0&  0&  0&  1&  0&   3\\
AN          &  0&  0&  0&  1&  0&  0&  0&  1&  0&  0&   2\\
PASJ        &  0&  0&  0&  0&  1&  0&  0&  0&  1&  0&   2\\
NuPhA       &  0&  0&  0&  0&  0&  2&  0&  0&  0&  0&   2\\
PhST        &  0&  0&  0&  0&  0&  0&  0&  0&  2&  0&   2\\
Others$^{\rm b}$ &  2&  0&  0&  0&  4&  0&  1&  0&  2&  2&  11\\
Sum         &161&161&176&199&208&217&265&308&262&269&2226\\
\hline
\end{tabular}
\medskip\\
\end{center}
$^{\rm a}$The only one paper published in $A\&A~ Supplements$
  in 2000 is included here \\
$^{\rm b}$Journals with only one paper during $2000-2009$ \\
\end{table*}

\begin{table*} 
\begin{center}
\caption{Number of Papers Published by Using the VLT During $2000-2009$}\label{tab5}
\begin{tabular}{lrrrrrrrrrrr}
\hline\hline
{Journal} & {2000} & {2001} & {2002}& {2003}
 & {2004} & {2005} & {2006} & {2007} & {2008} & {2009} & {Sum}\\
\hline
A\&A        & 35& 56& 86&129&177&176&204&230&228&210&1531\\
ApJ         & 11& 23& 42& 66& 81& 78& 95&106&103&115& 720\\
MNRAS       &  1&  5& 11& 26& 33& 51& 59& 80& 93& 67& 426\\
AJ          &  2&  7&  8& 20& 16& 14& 18& 14& 13& 15& 127\\
ApJS        &  0&  0&  2&  0&  5&  5&  3& 21&  8&  5&  49\\
Nature      &  1&  6&  2&  4&  4&  7&  9&  1&  6&  6&  46\\
Icarus      &  0&  0&  0&  3&  3&  5&  1&  7&  4&  3&  26\\
Science     &  0&  1&  2&  3&  0&  0&  2&  5&  4&  1&  18\\
AN          &  0&  2&  1&  2&  0&  0&  1&  2&  3&  5&  16\\
PASP        &  0&  0&  0&  0&  1&  6&  2&  1&  1&  4&  15\\
JGR         &  0&  0&  0&  0&  0&  0&  3&  2&  1&  0&   6\\
NewAR       &  0&  4&  0&  0&  1&  0&  0&  0&  0&  0&   5\\
NewA        &  0&  0&  1&  0&  0&  1&  0&  0&  0&  2&   4\\
EM\&P       &  0&  0&  3&  0&  0&  0&  0&  0&  0&  0&   3\\
P\&SS       &  0&  0&  0&  0&  0&  1&  0&  0&  1&  1&   3\\
AdSpR       &  0&  0&  0&  0&  2&  0&  0&  0&  0&  0&   2\\
PhRvL       &  0&  0&  0&  0&  1&  0&  1&  0&  0&  0&   2\\
ARA\&A      &  0&  0&  0&  0&  0&  0&  0&  0&  1&  0&   1\\
Others$^{\rm a}$ &  0&  1&  0&  0&  2&  0&  0&  1&  0&  2&   6\\
Sum         & 50&105&158&253&326&344&398&470&466&436&3006\\
\hline
\end{tabular}
\medskip\\
\end{center}
$^{\rm a}$Journals with only one paper during $2000-2009$ \\
\end{table*}

\begin{table*} 
\begin{center}
\caption{Number of Papers Published by Using the Gemini Telescopes During $2000-2009$}\label{tab6}
\begin{tabular}{lrrrrrrrrrrr}
\hline\hline
{Journal} & {2000} & {2001} & {2002}& {2003}
 & {2004} & {2005} & {2006} & {2007} & {2008} & {2009} & {Sum}\\
\hline
ApJ         &  7& 12& 18& 18& 33& 46& 56& 89& 71& 89& 439\\
MNRAS       &  2&  6&  4&  5& 10& 13& 27& 33& 29& 38& 167\\
AJ          &  1&  4&  9& 14& 10& 19& 22& 20& 22& 21& 142\\
A\&A        &  2&  6&  5&  5&  7&  5& 12& 11& 22& 21&  96\\
Icarus      &  2&  2&  3&  0&  1&  3&  1&  4&  5&  2&  23\\
Nature      &  1&  0&  0&  1&  1&  5&  6&  1&  3&  3&  21\\
ApJS        &  0&  0&  0&  1&  2&  1&  2&  4&  3&  4&  17\\
PASP        &  0&  3&  4&  1&  4&  1&  1&  1&  1&  0&  16\\
Science     &  1&  0&  0&  1&  1&  3&  3&  0&  5&  1&  15\\
NewAR       &  0&  0&  0&  0&  0&  0&  5&  0&  0&  0&   5\\
Ap\&SS      &  0&  0&  3&  0&  0&  0&  0&  1&  0&  1&   5\\
JOSAA       &  0&  2&  2&  0&  0&  1&  0&  0&  0&  0&   5\\
AN          &  0&  0&  0&  0&  1&  1&  0&  0&  0&  1&   3\\
RMxAA       &  0&  0&  0&  2&  0&  0&  0&  0&  1&  0&   3\\
PASA        &  0&  0&  0&  0&  1&  0&  0&  0&  0&  1&   2\\
ApOpt       &  0&  0&  0&  2&  0&  0&  0&  0&  0&  0&   2\\
P\&SS       &  0&  0&  1&  0&  0&  0&  0&  0&  0&  1&   2\\
Others$^{\rm a}$ &  0&  0&  2&  0&  0&  1&  1&  2&  0&  2&   8\\
Sum         & 16& 35& 51& 50& 71& 99&136&166&162&185& 971\\
\hline
\end{tabular}
\end{center}
$^{\rm a}$Journals with only one paper during $2000-2009$ \\
\end{table*}

\begin{table*} 
\begin{center}
\caption{Number of Papers Published by Using the Subaru Telescope During $2000-2009$}\label{tab7}
\begin{tabular}{lrrrrrrrrrrr}
\hline\hline
{Journal} & {2000} & {2001} & {2002}& {2003}
 & {2004} & {2005} & {2006} & {2007} & {2008} & {2009} & {Sum}\\
\hline
ApJ         &  0& 11& 22& 25& 31& 30& 43& 28& 37& 65& 292\\
PASJ        & 15& 10& 21&  9& 11& 12&  8& 10& 15& 13& 124\\
AJ          &  1&  1&  1&  9& 13&  6&  8&  6&  5&  5&  55\\
MNRAS       &  0&  0&  3&  4&  5&  4& 10&  9&  9&  4&  48\\
A\&A        &  1&  0&  3&  0&  4&  5&  2&  5& 11& 13&  44\\
ApJS        &  0&  0&  0&  1&  1&  0&  2& 33&  6&  1&  44\\
Nature      &  0&  0&  0&  1&  1&  2&  2&  1&  1&  1&   9\\
Science     &  0&  1&  0&  0&  1&  2&  0&  0&  2&  0&   6\\
Icarus      &  0&  0&  0&  0&  0&  0&  0&  1&  1&  2&   4\\
PASP        &  0&  0&  0&  0&  0&  0&  1&  0&  0&  2&   3\\
JGR         &  0&  0&  0&  0&  0&  0&  3&  0&  0&  0&   3\\
AN          &  0&  0&  0&  0&  0&  1&  0&  0&  1&  0&   2\\
P\&SS       &  0&  0&  0&  0&  0&  0&  0&  2&  0&  0&   2\\
Others$^{\rm a}$&  0&  0&  0&  1&  1&  1&  2&  0&  0&  1&   6\\
Sum         & 17& 23& 50& 50& 68& 63& 81& 95& 88&107& 642\\
\hline
\end{tabular}
\end{center}
$^{\rm a}$Journals with only one paper during $2000-2009$ \\
\end{table*}

\begin{table*} 
\begin{center}
\caption{Number of Papers Published by Using the HET During $2000-2009$}\label{tab8}
\begin{tabular}{lrrrrrrrrrrr}
\hline\hline
{Journal} & {2000} & {2001} & {2002}& {2003}
 & {2004} & {2005} & {2006} & {2007} & {2008} & {2009} & {Sum}\\
\hline
ApJ         &  3&  3&  2&  4& 11&  3& 10& 10&  6&  2&  54\\
AJ          &  4&  7&  5&  9&  6&  3&  7&  4&  4&  0&  49\\
A\&A        &  1&  1&  2&  3&  0&  1&  1&  2&  1&  0&  12\\
PASP        &  1&  2&  1&  1&  0&  0&  0&  0&  0&  1&   6\\
MNRAS       &  0&  0&  0&  2&  1&  0&  1&  1&  0&  0&   5\\
ApJS        &  0&  0&  0&  0&  1&  0&  0&  1&  1&  2&   5\\
Sum         &  9& 13& 10& 19& 19&  7& 19& 18& 12&  5& 131\\
\hline
\end{tabular}
\end{center}
\end{table*}



\begin{thebibliography}{} 

\bibitem[Abt (1980)]{abt80}Abt, H. A. 1980, PASP, 92, 249

\bibitem[Abt (1995)]{abt95}Abt, H. A. 1995, ApJ, 455, 407

\bibitem[Abt (1998)]{abt98}Abt, H. A. 1998a, PASP, 110, 210

\bibitem[Abt (2000)]{abt00}Abt, H. A. 2000, PASP, 112, 1417

\bibitem[Abt (2010)]{abt10}Abt, H. A. 2010, PASP, 122, 955

\bibitem[Ahn et al. (2008)]{ahn08}Ahn, S.-H., Park, B.-G., Kim, Y.-S.,
  Chun, M.-Y., Kim, H.-I., Sung, H.-I., Lee, D.-W., \& Kim, S. C.
  2008, PKAS, 23, 123 (Erratum: 2010, PKAS, 25, 51)

\bibitem[Apai et al. (2010)]{apai10}Apai, D., Lagerstrom, J., Reid, I. N., 
  Levay, K. L., Fraser, E., Nota, A., \& Henneken, E. 2010, PASP, 122, 808

\bibitem[Barth et al. (2003)]{barth03} Barth, A. J., Sari, R., Cohen, M. H., Goodrich, R. W.,
  Price, P. A., Fox, D. W., Bloom, J. S., Soderberg, A. M., Kulkarni, S. R.
  2003, ApJ, 584, L47

\bibitem[Benn \& S{\'a}nchez (2001)]{benn01}Benn, C. R., \& S{\'a}nchez, S. F.
   2001, PASP, 113, 385

\bibitem[Bracher (1999)]{bracher99} Bracher, K. 1999, AJ, 117, 12

\bibitem[Crabtree (2008)]{crabtree08} Crabtree, D. 2008, SPIE, 7016, 70161A

\bibitem[Crabtree (2011)]{crabtree11} Crabtree, D. 2011, AAS, 217, 15719

\bibitem[Cuby et al. (2007)]{cuby07} Cuby, J.-G., et al. 2007,
  A preparatory document to the 2007 CFHT Users' Meeting
  (http://www.cfht.hawaii.edu/ \\ UM2007/UM07\_Prep.pdf)

\bibitem[Da Rocha et al. (2002)]{darocha02}
  Da Rocha, C., Mendes de Oliveira, C., Bolte, M., Ziegler, B. L., \& Puzia, T. H.
  2002, AJ, 123, 690

\bibitem[Davoust \& Schmadel (1987)]{davoust87}Davoust, E., \& Schmadel, L. D.
  1987, PASP, 99, 700

\bibitem[De Breuck et al. (2001)]{deBreuck01}De Breuck, C. 
  et al. 2001, AJ, 121, 1241

\bibitem[Dessauges-Zavadsky et al. (2002)]{dessaugeszavadsky02}
  Dessauges-Zavadsky, M., Prochaska, J. X., \& D'Odorico, S.
  2002, A\&A, 391, 801

\bibitem[Drory et al. (2001)]{drory01} Drory, N. 
  et al. 2001, ApJ, 562, L111

\bibitem[Fischer et al. (2005)]{fischer05} Fischer, D. A. et al. 2005, ApJ, 620, 481

\bibitem[Frey \& Osterloh (2010)]{frey10} Frey, B. S., \& Osterloh, M. 2010, 
  Nature, 465, 871

\bibitem[Frogel (2010)]{frogel10}Frogel, J. A. 2010, PASP, 122, 1214 

\bibitem[Grothkopf et al. (2005)]{grothkopf05} Grothkopf, U., Leibundgut, B.,
  Macchetto, D., Madrid, J. P., \& Leitherer, C. 2005, The Messenger, 119, 45

\bibitem[H\"oflich et al. (2004)]{hoeflich04} H\"oflich, P. et al. 2004, ApJ, 617, 1258

\bibitem[Hu et al. (2002)]{hu02} Hu, E. M. 
  et al. 2002, ApJ, 568, L75

\bibitem[Leverington (1996)]{leverington96}Leverington, D. 1996,
  Q. J. R. astr. Soc., 37, 643

\bibitem[Leverington (1997a)]{leverington97a}Leverington, D. 1997a,
  Nature, 385, 196

\bibitem[Leverington (1997b)]{leverington97b}Leverington, D. 1997b,
  Nature, 387, 12

\bibitem[Meylan et al. (2004)]{meylan04}Meylan, G., Madrid, J. P., \& 
  Macchetto, D. 2004, PASP, 116, 790 

\bibitem[Metcalfe (2005)]{metcalfe05}Metcalfe, T. S. 2005, BAAS, 37, 555

\bibitem[Pettini \& Bowen (2001)]{pettini01} Pettini, M., \& Bowen, D. V.
  2001, ApJ, 560, 41

\bibitem[Ringwald et al. (2003)]{ringwald03}Ringwald, F. A., Culver, J. M.,
   Lovell, R. L., Kays, S. A., \& Torres, Y. V. 2003, BAAS, 35, 1063

\bibitem[Schaye et al. (2000)]{schaye00}Schaye, J., Rauch, M., Sargent, W. L. W., 
  \& Kim, T.-S. 2000, ApJ, 541, L1

\bibitem[Schulman et al. (1997)]{schulman97}Schulman, E., French, J. C.,
  Powell, A. L., Eichhorn, G., Kurtz, M. J., \& Murray, S. S.
  1997, PASP, 109, 1278

\bibitem[Stanek (2008)]{stanek08} Stanek, K. Z. 2008, (arXiv:0809.0692)

\bibitem[Trimble (1995)]{trimble95} Trimble, V. 1995, PASP, 107, 977

\bibitem[Trimble (1996)]{trimble96} Trimble, V. 1996, Scientometrics, 36, 237

\bibitem[Trimble (2009)]{trimble09} Trimble, V. 2009, Exp. Astron., 26, 133

\bibitem[Trimble, Zaich, \& Bosler (2005)]{trimble05} Trimble, V., Zaich, P.,
  \& Bosler, T. 2005, PASP, 117, 111

\bibitem[Trimble \& Ceja (2007)]{trimble07} Trimble, V., \& Ceja, J. A.
   2007, AN, 328, 983

\bibitem[Trimble \& Ceja (2008)]{trimble08} Trimble, V., \& Ceja, J. A.
   2008, AN, 329, 632

\bibitem[Veillet (2011)]{veillet11} Veillet, C. 2011, private communication

\bibitem[Venn et al. (2001)]{venn01} 
Venn, K. A. 
  et al. 2001, ApJ, 547, 765

\bibitem[Vreeswijk et al. (2004)]{vreeswijk04} Vreeswijk, P. M. 
  et al. 2004, A\&A, 419, 927

\bibitem[White (2007)]{white07}White, S. D. M. 2007, RPPh, 70, 883

\bibitem[Zheng et al. (2000)]{zheng00} Zheng, W. 
  et al. 2000, AJ, 120, 1607

\end{thebibliography}
\end{document}